\documentclass[journal=jpcbfk,manuscript=article]{achemso}
\usepackage[sort&compress,numbers]{natbib}
\usepackage{amsmath}
\usepackage[normalem]{ulem}
\usepackage{color}
\usepackage{graphicx}
\usepackage{dcolumn}
\usepackage{bm}
\usepackage{caption}
\usepackage{float}
\usepackage{setspace}
\usepackage{xkeyval}
\usepackage{graphicx}
\usepackage{epstopdf}

\usepackage[version=3]{mhchem} 
\setkeys{acs}{etalmode=truncate,maxauthors=10}

\author{Anthony J. Leggett}
\affiliation[University of Illinois at Urbana-Champaign]{Dept. of Physics, University of Illinois at Urbana-Champaign, 1110 W. Green St, Urbana, IL, USA; Tel: 217.333.2077 }
\email{aleggett@illinois.edu}
\author{Dervis C. Vural}
\affiliation[Harvard University]{Applied Physics, SEAS, Harvard University, 29 Oxford St, Cambridge, MA, USA; Tel: 617.401.5659}
\email{dcvural@seas.harvard.edu}
\title{The ``Tunneling Two-Level Systems'' Model of the Low-Temperature Properties of Glasses:  Are "Smoking-Gun" Tests Possible?}

\begin{document}
\begin{abstract}
  Following a brief review of the ``two-level (tunneling) systems'' model of the low-temperature properties of amorphous solids ("glasses"), we ask whether it is in fact the unique explanation of these properties as is usually assumed, concluding that this is not necessarily the case. We point out that (a) one specific form of the model is already experimentally refuted, and (b) that a definitive test of the model in its most general form, while not yet carried out, would appear to be now experimentally feasible.
\end{abstract}

\section{Introduction}
Structurally amorphous materials (``glasses'') constitute a substantial fraction of all terrestrial matter, yet our overall understanding of their behavior is not at all comparable to that which we have for crystalline matter. While both the transition to the glassy state and the thermodynamic and response behavior at ambient temperatures have some universal features which have been the subject of a huge literature (see e.g.  refs. \cite{wolynesLub,berthier} respectively), a particularly intriguing problem is posed by the behavior of glasses below about 1K. Ever since the pioneering experiments of Pohl and co-workers in the early 70's\cite{zeller}, it has been recognized that in this regime the properties of glasses are not only qualitatively different from those of crystalline solids but show a remarkable degree of universality. For example, almost without exception the specific heat of an arbitrary (insulating) amorphous solid well below 1K is approximately linear in $T$, the thermal conductivity approximately quadratic in $T$ and the ultrasonic behavior consistent with a $Q$-factor which at zero temperature is independent of frequency and surprisingly large (more on this below).\\

Very soon after the original experiments\cite{zeller} a plausible model to explain them was published independently by Phillips\cite{phillips1972} and by Anderson,  
Halperin and Varma\cite{anderson} .  This model,  which has become known as the ``tunneling two-level system'' (TTLS) model, postulates that because of the structurally amorphous nature of the system there exist some entities (single atoms,  groups of atoms or in some cases even single electrons) which have available two nearly degenerate configurations and can tunnel between them (a more quantitative description is given in the next section).   With a plausible choice of the distribution of the relevant parameters, the TTLS model can account for the qualitatively universal features noted above;   furthermore,  by analogy with other well-known examples of ``two-state'' systems in e.g. atomic physics and NMR it naturally predicts various nonlinear phenomena such as acoustic saturation and echoes, all of which have been at least qualitatively verified by experiment\cite{phillips1987}. The original model has undergone considerable elaboration over the last 40 years;  in particular,   an interesting series of papers\cite{lubchenko} by Peter Wolynes and his collaborators has attempted to explain the generic existence of the postulated two-level systems as a natural consequence of processes occurring in the glass transition.   These successes have persuaded the overwhelming majority of the relevant community that the TTLS model is the unique explanation of the low-temperature properties of glasses.    \\

Viewed from the above perspective,   the aim of the present note may seem rather quixotic: to ask whether the TTLS explanation is indeed as unique as it is usually taken to be,   and to try to suggest definitive ways of testing the model against a particular (loosely defined) class of alternative hypotheses.   Obviously,   to do this it is necessary to define exactly what we mean by the ``TTLS model'' (and in particular what it excludes), and this will be done in section 2.   In section 3 we sketch some reasons for skepticism about the model,   and introduce a (very generic) class of alternative hypotheses,   and in section 4 we propose an experiment which we believe should discriminate unambiguously between this class and the TTLS model. Throughout we confine ourselves to insulating glasses and concentrate on the linear ultrasonic properties, which we believe are among the most unambiguously predicted consequences of the model.   

\section{Definition of the TTLS model; some simple consequences}
There exist in the literature a number of good accounts of the TTLS model and its most important experimental predictions\cite{phillips1987,enss}. For present purposes a sufficient definition runs roughly as follows: For the purpose of calculating the low-energy states of glasses (those relevant to the equilibrium and near-equilibrium properties well below 1K) an adequate effective Hamiltonian is of the form

\begin{equation}
\widehat{H}=\widehat{H}_{ph}+\widehat{H}_{TLS}+\widehat{H}_{int}
\end{equation}

Here $\widehat{H}_{ph}$ is the "phonon" Hamiltonian, given in terms of phonon annihilation and
creation operators $a^{+}_{i}$, $a_i$ by

\begin{equation}
   \widehat{H}_{ph}=\sum\limits_{k \alpha} \hbar \omega_{k\alpha} a^{+}_{k\alpha} a_{k\alpha}, \indent \indent \omega_{k\alpha} = c_\alpha |k|
\end{equation}

where the sum over k has an upper cutoff $k_{max}$ such that $\hbar c_i k_{max} \gg k_B \times 1K$, and where the sum over $\alpha$ runs over one longitudinal ($l$) and two transverse ($t$) branches; the sound velocities $c_l$, $c_t$ are constant, reflecting the average isotropy of the glass at sufficiently large length scales. $\widehat{H}_{TLS}$ is the part of the Hamiltonian attributable to the postulated two-level systems, and has the form
\begin{equation}
    \widehat{H}_{TLS}=\sum\limits_{i} E_i b_{i}^{+}  b_i
\end{equation}
where the operators $b_{i}^{+}$, $b_i$ are Pauli operators, i.e. they satisfy the (anti)commutation relations
\begin{equation}
     \{b_i, b_{i}^{+}\}=1, \indent \indent [b_i, b_{j}^{+}]= 0 \text{ for } i \neq j
\end{equation}
Thus the eigenvalues of the "occupation numbers" $n_i=b_{i}^{+} b_i$ are 1 and 0, and can be specified independently. Finally, the part of the Hamiltonian expressing the interaction between the phonons and the TLS is of the form
\begin{equation}
    \widehat{H}_{int}=\sum\limits_{\alpha \beta} \int \hat{e}_{\alpha \beta}({\bf r})  \hat{T}_{\alpha \beta}({\bf r})d{\bf r}
\end{equation}
where $\hat{e}_{\alpha \beta}= \frac{1}{2}(\frac{\partial u_\alpha}{\partial x_\beta} + \frac{\partial u_\beta}{\partial x_\alpha})$ is the phonon strain operator ($u_\alpha =$
mean "background material" displacement) and the stress operator $T_{\alpha \beta}$ is linear in the ``spin'' operators $\hat{\sigma}^{(i)}$ of the TLS: 
\begin{equation}
    T_{\alpha \beta}({\bf r})= \sum\limits_{i} g_{\alpha \beta \gamma}^{(i)}  \hat{\sigma}_\gamma^{i}\delta({\bf r}-{\bf r}_i)
\end{equation}
where ${\bf r}_i$ is the position of the TLS $i$\\

Several points should be noted about the model described by eqns. (1-6): First, it excludes what one might think are some obvious possibilities; most obviously, that the non-phonon part of the Hamiltonian cannot be written in the simple additive form (3), but also (e.g.) that the stress tensor involves combinations of the $\hat{\sigma}^{(i)}$'s higher than linear, e.g. $K_{ij} \hat{\sigma}^{(i)} \hat{\sigma}^{(j)}$. Secondly, at the present stage it is really only a generic ``two-level'' model; the ``tunneling''  aspect enters only through the choice of the parameters $E_i$ and $g_{\alpha \beta \gamma}^{(i)}$ (see below). Thirdly-and this is crucial to the subsequent discussion-it is implicit that the coupling term (5) is small enough that in the calculation of any given physical property it is adequate to confine oneself to the lowest order in $\widehat{H}_{int}$ which gives ``physically sensible'' results (e.g. for the specific heat, zeroth order and for the ultrasonic attenuation, first order). Finally, it should be noted that it is not necessarily assumed that the Hamiltonian (1) is the one which would be appropriate at a fully microscopic level; bearing in mind that we are interested only in ``low-energy'' phenomena, and hence by implication only in the behavior of phonons with wavelength $\lambda > hc/k_B T$ (which for $T\sim1K \text{ is } \sim1000 \buildrel _\circ \over {\mathrm{A}}$), we may include the possibility that (1) is itself the output of some nontrivial renormalization procedure. \\

Eqns. (1)-(6) may be regarded as the most general definition of the ``TLS'' (two-level system) model of glasses. However, to extract quantitative physical predictions from it one of course needs to specify the distribution of the parameters $E_i$ and $g_{\alpha \beta \gamma}^{(i)}$ (including any correlations between these two quantities). This is where the ``tunneling'' aspect comes in; in the original and simplest version of the model the two states in question are conceived as corresponding to two spatially separated positions of the center of mass of a group of particles, so that for as given TLS i, in a basis in which  these two positions are taken as the eigenstates of $\hat{\sigma}_{z}^{(i)}$, the Hamiltonian is
\begin{equation}
    \widehat{H}_i= \frac{1}{2}  \begin{pmatrix}
			   \epsilon_i  & \Delta_i \\
                		  \Delta_i   &-\epsilon_i
		        \end{pmatrix}
\end{equation}
with $\epsilon_i$ the offset between the potential energy in the two configurations and $\Delta_i$ the matrix element for tunneling between them. Then we evidently have
\begin{equation}
      E_i= (\epsilon_i^2+\Delta_i^2)^{1/2} 
\end{equation}
What is the distribution of the offsets $\epsilon_i$ and the tunneling matrix elements $\Delta_i$? As regards the former, it seems very reasonable to take it to be simply constant (uniform distribution; from a microscopic point of view there is nothing special about zero bias). As to the $\Delta_i$, the default assumption would seem to be that the principal dependence comes from the WKB exponent in the expression for the tunneling amplitude; then rather general arguments indicate that provided we are interested mainly in low-energy
states, this exponent can again be considered to be uniformly distributed. With these rather weak assumptions, the distribution of parameters in the two-dimensional ($\epsilon$, $\Delta$) space is
\begin{equation}
     \rho(\epsilon, \Delta)= \text{const.}/\Delta.
\end{equation}
with the constant of course material-dependent. This gives a  constant density
of states as a function of the total energy splitting E:
\begin{equation}
     \rho(E)= \bar{P}_0
\end{equation}
We also need to consider the distribution of the TLS-phonon coupling constants $g_{\alpha \beta \gamma}^{(i)}$. The usual assumption in the literature is that (a) any difference between the $l$ and $t$ components of $g$ lies only in the overall magnitude, not the form, and (b) strain of either kind mainly affects the offset $\epsilon$, so that any effect on the tunneling matrix element $\Delta$ may be neglected. If this is so, then in the energy representation for the Pauli matrices $\sigma^{(i)}$ (which we will use from now on unless explicitly otherwise stated) the matrix form of $g_{\alpha \beta \gamma}^{(i)}$ is identical to that of $\widehat{H}_{TLS}$ in the original (position) representation, i.e.
\begin{equation} 
      g_{\alpha \beta \gamma}^{(i)}=(g_{\alpha \beta}^{(i)}/E_i) (\epsilon_{i} \delta_{z \gamma}+\Delta_{i} \delta_{x\gamma})
\end{equation}
with the parameters $g_{\alpha \beta}^{(i)}$ having some random distribution uncorrelated with $\Delta_i, \epsilon_i$, e.g., a Gaussian with zero mean. When combined with (9), eqn. (11) says that the distribution of the off-diagonal (in the energy representation) terms in the $g^{(i)}$ is strongly peaked towards small values (an important qualitative feature of the model which was already recognized in the earliest papers). We will refer to the distribution of parameters given by eqns. (9) and (11) as the "canonical" distribution, and distinguish between the generic "TLS" model defined by eqns. (1-6) and the ``TTLS'' model defined by eqns. (1-6) \emph{plus} this distribution.

\section{What could be wrong with it? An alternative scenario}

Let us start with a semi-philosophical point which will probably be dismissed out of hand by most right-thinking physicists: the fact that the TTLS model appears to explain so adequately most of the properties of amorphous solids does not prove that it is correct! Technically, to argue that ``Theory T predicts experimental result(s) E; we see E; therefore theory T is correct'' would of course be to commit the logical fallacy known as ``affirming the consequent''. Of course, this fallacy is formally committed every day in the pages of physics journals, and generally people do not worry too much about it. Why not? We suspect because typically in these cases there is an unspoken extra premise: ``it is extremely unlikely that any theory other than $T$ would predict experimental results E'', which when combined with ``we see E'' indeed permits us to draw with high confidence the conclusion that $T$ is correct. The question then arises: in the case of the use of the TTLS model to explain the behavior of glasses, is the unspoken extra premise correct? We believe this is less obvious than it may seem. \\

First, let's consider the various kinds of nonlinear behavior in the ultrasonics (saturation, echoes, hole-burning ...), which we suspect are in most people's minds the most convincing evidence in favor of the model. Setting aside the rather complicated question of the degree of quantitative agreement of the data with the TTLS model, one may ask: is the prediction of the qualitative features, i.e. the mere existence of these phenomena, unique to the model? We believe that the answer is no, and indeed that it is probable that almost {\it any} model of the system energy levels and stress matrix elements other than the familiar harmonic-oscillator one will suffice to reproduce them; calculations on simple one-particle systems tend to confirm this prejudice. A simple "hand-waving" argument goes as follows: in quantum mechanics, to obtain a response to a near-monochromatic field which \emph{fails} to saturate, we need not just an infinite sequence of equidistantly spaced energy levels but also that the relevant matrix elements (in the case of ultrasound absorption,those of the stress tensor operator) between these levels increase sufficiently fast with energy. While we are used to the fact that these conditions are both adequately satisfied for the simple harmonic oscillator,they are really more like the exception than the rule. In other words, it may be that what is ``special'' about the non-phonon modes in glasses is not that they are well described as TTLS, but that they are {\it not} well described as simple harmonic oscillators! \\

Of course, there are other features of the experimental data whose prediction might at first sight seem unique to the TTLS model, such as time-dependent specific heats \cite{zhou} and the equality, up to a numerical factor, of ultrasonic absorption in the low-temperature resonant and high-temperature (low-frequency) relaxation regimes (see section 4); while our prejudice is that a more general scenario should be able to reproduce at least the qualitative aspects of these phenomena, it must be said at once that a quantitative calculation is at present lacking. \\

However, it is not always appreciated that to a large extent the same situation exists for the established model. Indeed, while it is probably true that just about all the existing experimental data is consistent with the generic "TLS" model, it can often be made so only by a choice of parameter distributions which violates the more restrictive assumptions (9) and (11) which were taken above to define the "TTLS" version. A typical example is the specific heat: the experimental dependence on temperature is actually not linear, but rather resembles a power law with exponent 1.2-1.3, and this behavior is clearly inconsistent with eqn.(10) and hence with the TTLS ansatz (9). More generally, once one abandons the defining ansatz (9) and (11) of the TTLS version, the TLS model becomes so "squishy" that a cynic might be forgiven for anticipating that it can be made consistent with just about any experimental data. \\

A rather different kind of motivation for challenging the uniqueness of the TTLS model as an explanation of the low-temperature properties of glasses lies in the striking quantitative universality of some of these properties, in particular the dimensionless (and surprisingly large, $\sim10^4$) $Q$-factor which describes the transverse ultrasonic absorption and frequency shift in the MHz-GHz range\cite{berret,pohl}. In the TTLS model, where the relevant expression is a product of four independent factors, this universality can be attributed only to a mind-boggling degree of coincidence; more generally, what it seems to suggest is that the low-temperature, long-wavelength properties of glasses emerge as a result of some rather nontrivial renormalization process which iterates to a fixed point. An initial attempt to implement such a renormalization scheme has been made in ref\cite{vural}. where it has been shown that with two unproved but plausible generic ans$\ddot{a}$tze it is possible to reproduce something like the observed value of Q (cf. also ref. \cite{burin}). Of course, this consideration does not in itself imply that the final Hamiltonian which emerges from the renormalization process does not itself possess the ``TLS'' structure described by eqns. (1-6), but from an intuitive point of view this seems rather unlikely. \\

With this motivation, let us consider, as a counterpoint to the TTLS model, a scenario \cite{vural} which is about as far from it as possible without describing simply a collection of harmonic oscillators, namely one in which while the phonon contribution to the total Hamiltonian is still given by eqn. (2), the "non-phonon" term $\widehat{H}_{TLS}$ (eqn. (3)) and the coupling term (eqn. (4)) are replaced by random matrices in the many-body Hilbert space with some appropriately specified statistical properties. Needless to say, we do not necessarily expect that this model will be the correct one-the truth may well lie somewhere between it and the TLS version-but it serves as a convenient point of comparison.\\

We would like to emphasize strongly that the conjecture made in this paper is \emph{not} that two-level systems in amorphous solids never exist. Indeed,there are a few systems in which their existence may be established rather directly by experiment. A particularly striking example comes from the beautiful experiments of the Bordeaux group\cite{boiron1999spectral} on the spectroscopy of single terrylene molecules in polyethylene (PET), in particular from the ``spectral trail'' experiments, which give rather direct evidence that the that in a substantial fraction of the molecules studied ($\sim40\%$) the part of the environment which gives rise to the shifts in the resonance frequency jumps between two (and only two) discrete configurations. However, what is interesting is that to the extent that one takes these experiments as definitive evidence for the presence of TLS in bulk amorphous PET, by the same token one will have to take the similar experiment of ref \cite{eremchev2011low} as equally definitive evidence for their \emph{absence} in amorphous solid toluene, and that consideration would suggest that it would be extremely interesting to investigate whether the sub-degree thermal and acoustic behavior of toluene is that of a typical amorphous solid; while a negative answer would strengthen the case for the TLS model, a positive one would definitively refute it.\\

Let us mention a few more systems in which either theory or experiment or both provides (or has been thought to provide) strong arguments for the presence of TLS. (1) It has long been believed that the behavior of KBr-KCN mixtures can be satisfactorily explained\cite{grannan1988low} in terms of TLS associated with the two possible orientations of the KCN complex. (2) In the disordered oxide barriers which are nowadays often used in Josephson junctions,there is rather direct evidence there exist TLS carrying electric dipole moments,which moreover can be tuned by the application of external strain \cite{grabovskij2012strain}. what is interesting is that rather frequently these systems turn out to have a value of $Q^{-1}$ considerably smaller than the "universal" figure of $\sim 3 \times 10^{-4}$. Our conjecture, therefore, would be that for these comparatively rare cases the renormalization process discussed in ref. \cite{vural} is ineffective; while for the majority of cases it takes place and leads not only to the universal (maximum allowed) value of $Q^{-1}$ but also to a structure of the output many-body levels which is in general of non-TLS form. (3) Finally, it may be claimed that the isotope effects observed\cite{nagel2004novel} in (natural or deuterated) glycerol are definitive proof of the TLS model. This brings us back to the point raised at the beginning of this section: While we of course  agree that the TLS model gives a natural and elegant explanation\cite{wurger2004dephasing} of the data, the interesting question is whether it is unique in doing so, and this may be regarded as a special case of the issue raised above concerning nonlinear effects more generally.\\

To repeat, in this paper we are not disputing that there are some amorphous solids in which TLS exist; we are not even necessarily disputing that TLS may exist, at some level, in all amorphous solids. What we \emph{are} disputing is the claim that the TLS model, in the precise sense in which we have defined it in the last section, is the unique and universal explanation of the behavior, in particular the thermal and ultrasonic behavior, of amorphous solids below 1K.

\section{Ultrasonic absorption -a smoking gun?}
The classic work on ultrasound propagation within the TTLS model, with the distribution of parameters given by the "canonical" form (eqns. (9) and (11)), is the paper of J$\ddot{a}$ckle \cite{jaeckle}. Here we briefly review the main conclusions, confining ourselves to the small-amplitude (linear) regime. To facilitate the discussion it is convenient to define the quantity $\tau_c(T)$, the characteristic relaxation time of a symmetric ($E_j=\Delta_j$) TLS with splitting $E$ equal to $k_B T$; this is given by the appropriate special case
of eqn. (7) of ref \cite{jaeckle} and may be verified to be proportional to $T^3$. A closely similar discussion may be given of dielectric-loss experiments (see ref. \cite{enss}). \\

Within the model there are two mechanisms for ultrasound absorption which operate in parallel. The first is resonant absorption; the relevant expression follows directly from standard ``golden-rule'' perturbation theory in $\widehat{H}_{int}$, and is given by \cite{enss}
\begin{equation}
Q_{res}^{-1}(\omega)=\pi C \tanh(\hbar \omega/2k_B T), \indent \indent C \equiv \bar{P}_0 \gamma^2/\rho c^2
\end{equation}
where $\bar{P}_0$ is the (constant) TTLS density of states (eqn. (10)), and $\gamma$ is the rms value of the coupling constant $g_{\alpha \beta}$. (Actually, the quantities Q, C , c and $\gamma$ should each have a suffix $\alpha=l, t$ denoting the polarization of the sound mode in question; we omit this to avoid cluttering up the formulae). \\

The second process, "relaxation" absorption, is a little more subtle: The phonon modulates the energy splitting of the TLS, and thereby throws their occupations out of thermal equilibrium; the ensuing relaxation to equilibrium extracts energy from the sound wave and thus leads to damping. The relevant formula (equivalent to eqns. (21) and (22) of ref. \cite{jaeckle}) is
\begin{equation}
    Q_{rel}^{-1}(\omega)= \text{const. } \sum\limits_{j} \frac{(-\partial n_0/\partial E_j) \omega \tau_j}{1+\omega^2 \tau_j^2} \gamma_j^2 (e_j/E_j)^2
\end{equation}
In the present context we will be interested primarily in the limit $\omega\ll\tau_c^{-1}(T)$ (but need to note that in the opposite limit the absorption is smaller by a factor o$(1/\omega \tau_c))$. In that limit, using the perturbation-theory formula for $\tau_j^{-1} (\propto(\Delta_j/E_j)^2)$ and the canonical distribution (9), we find, remarkably, that the distribution of $\tau^{-1}$ is, apart from a numerical constant, simply proportional to $\tau$, and this then implies the simple result
\begin{equation}
    Q_{rel}^{-1}= \frac{\pi}{2} C
\end{equation}
Comparing eqn. (14) with eqn. (11), we see that the relaxation contribution to the absorption (inverse Q-factor) in the low-frequency, high-temperature regime is predicted to be exactly half of its value in the high-frequency, low-temperature ($\hbar \omega\gg k_B T$) regime! \\

Let's now turn from the absorption to the shift of the ultrasound frequency due to interaction with the TLS. While the absolute value of this shift is of course not experimentally accessible, its temperature-dependence is, and as shown in ref. \cite{enss} a Kramers-Kronig analysis applied to the above results unambiguously predicts that for fixed $\omega$ one should find both for $\omega \tau_c \gg 1$ ($T\ll T_0$ where $T_0$ is the temperature at which $\omega \tau_c\sim1$) and for $\omega \tau_c \ll 1 (T\gg T_0)$ the relative velocity shift should be given by
\begin{equation}
      \delta c/c=A \ln(T/T_0)
\end{equation}
with A equal to C (eqn. (12)) on the low-temperature side and to -2C on the high-temperature side. This is {\it not} the behavior seen experimentally; while $\delta c/c$ indeed passes through a maximum for $T \sim T_0$, the (negative) slope for $T>T_0$ has a magnitude approximately equal to its (positive) value for $T<T_0$.\cite{classen} Ref. \cite{classen} notes that this and some other discrepancies with the TLS predictions cannot be fixed by "minor modifications" of the TTLS model. \\

While the above discrepancy may be a definitive refutation (at least for the relevant experimental systems, vitreous silica and BK7) of the TLS model when supplemented with the ``canonical'' parameter distribution (9), (i.e. of the TTLS model) is it a refutation of the TLS model as such? Since one needs to fit the experimentally observed frequency -as well as temperature- dependence, this is not immediately clear. So we may need to look elsewhere for our "smoking gun". \\

Consider then the absorption in the high-frequency, low-temperature regime ($\hbar \omega\gg k_B T, \hbar \tau_c^{-1}$). In this regime the only non-negligible contribution to absorption should be from the resonance mechanism, and a little thought shows that for quite general forms of the distribution of TLS splittings $\rho(E)$ it should have the general form
\begin{equation}
    Q^{-1}(\omega)=\text{ const. } (\rho (\hbar \omega) \tanh(\hbar \omega/2k_B T))
\end{equation}
(where the constant involves the $\gamma$'s, etc.). Thus, if we fix $\omega$ and vary $T$, we should predict quite independently of all the unknown parameters
\begin{equation}
      Q^{-1}(T)=\text{ const. } \tanh(x/2) \text{    } (x \equiv \hbar \omega / k_B T)
\end{equation}
This really is a smoking gun---violations of this prediction cannot be fixed by making ad hoc adjustments to the TLS model. Conversely, if the behavior (16) is indeed found, this would (pace the ``affirmation of the consequent'' objection!) be convincing evidence in favor of the model. \\

What is the current experimental situation? Many reviews give the impression that the TTLS prediction (17) is well verified. However, to the best of our knowledge, until very recently there was only one experiment \cite{golding} which evaded the low-frequency limit $x \equiv (\hbar \omega / k_BT) \ll 1$, and even that went up only to $x \sim 1.2$, a point at which the correction to the high-temperature limit formula is only about 20$\%$ (cf. fig. 2 of ref. \cite{golding}). So one may reasonably ask whether the data is equally consistent with an alternative formula, e.g.one that would follow from the alternative scenario sketched at the end of section 3. Actually, without further specification of this scenario the question is ill-defined, since unlike in the TLS scenario the stress matrix element $T_{mn}$ may depend on $E_m$ as well as the difference $E_m-E_n$. If for simplicity we postulate that $T_{mn}$ (or rather its statistical distribution) depends only on the difference $E_m-E_n$, then for fixed $\omega$ we recover a formula similar to (17), namely
\begin{equation}
	Q^{-1}(T)= \text{ const. } (1-\text{exp}-x)   \indent \indent        (x \equiv \hbar \omega/ k_B T) 
\end{equation}
In fig. (2) of ref. \cite{vural} is plotted a comparison of both (17) and (18) with the data of ref. \cite{golding}; while the former may seem to represent it slightly better, the difference would seem to lie within the presumed error bars. (In passing, we note that appreciably better agreement for both the velocity shift and the thermal conductivity is obtained for the "alternative" scenario than for the TTLS model \cite{vural}.) \\

In the last few years a number of experiments, primarily motivated by interest in designing high-Q superconducting circuits, have been performed on the {\it dielectric} loss of various amorphous materials; some of these have operated in the regime $\hbar \omega/k_B T > 1$ (see e.g. ref. \cite{kumar}). Unfortunately, while many of these papers claim evidence for (tunneling) two-level systems, none has to our knowledge tested explicitly for the characteristic TLS temperature-dependence (17) over a range where the discrepancy with (18) would become visible. We believe that such a test, preferably on a material with the "universal" value $\sim 3 \times 10^{-4}$ of $Q^{-1}$ (cf. end of section 3) is now feasible and would be a definitive test of the TLS model for such a system. 
\begin{acknowledgement}
It is a pleasure to dedicate this paper to Peter Wolynes and to wish him many more happy and fruitful years of research on glasses and his many other areas of interest. We thank our collaborators Pragya Shukla and Di Zhou for many discussions of the issues raised here. This work was supported by the National Science Foundation under award no. NSF- DMR09-06921.
\end{acknowledgement}
\bibliography{Wolynes_manuscript}
\pagebreak
\begin{figure}[h!]
\fbox{\includegraphics[width=2in, height=2in]{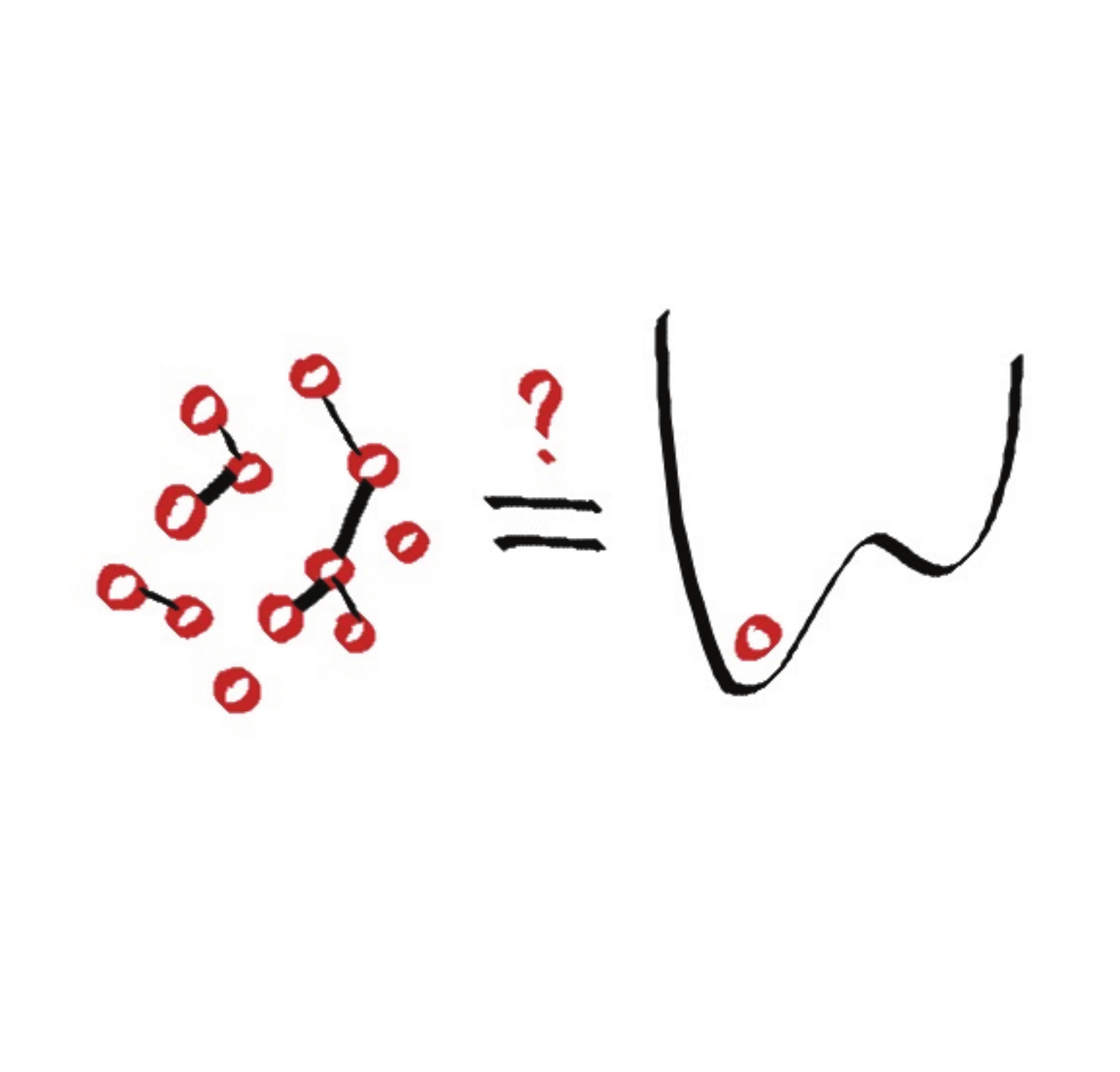}}
\end{figure}

\end{document}